# Online VNF Placement and Chaining for Value-added Services in Content Delivery Networks

Narjes Tahghigh Jahromi, Somayeh Kianpisheh, Roch H. Glitho

Concordia University, Montreal, QC, Canada

{n_tahghi, s_kianpi, glitho}@encs.concordia.ca

***Abstract*— Value-added Services (VASs) (e.g. dynamic site acceleration, media management) play a critical role in Content Delivery Networks (CDNs). Network Functions Virtualization (NFV) enables the agile provisioning of VASs. In NFV settings, VASs are provisioned as ordered sets of Virtual Network Functions (VNFs), forming VNF-Forwarding Graphs (VNF-FG) which are deployed in the CDN infrastructure. The CDN VAS VNF-FGs have a specific characteristic: they have one end-point (corresponding to the content server) that is unknown, prior to their placement. The proposals for CDN VAS VNF-FG placement, so far, have only considered offline placement, where the VNF-FGs are placed before end-user traffic steers into the network. However, in concrete cases, a change in service usage patterns might occur, a situation that could require a VNF-FG placement in an online manner. This paper tackles the problem of online VNF-FG placement for VASs in CDNs, taking into account the eventual reuses and migrations of already-deployed VNFs. A cost model is considered, including multiple costs; i.e. new VNF instantiations, migration, hosting and routing costs. The objective is to optimally place the VNF-FGs such that total reconfiguration costs are minimized while QoS is satisfied. An Integer Linear Programming (ILP) formulation is provided and evaluated in a small-scale scenario.**

***Keywords—Network Functions Virtualization (NFV), VNF placement, VNF-FG, Service Function Chaining (SFC), Content Delivery Network (CDN), Value-added Services (VAS).***

## I. Introduction

CDNs have gained immense popularity for their efficient delivery of content to a large number of geographically distributed end-users. CDNs are designed as an overlay network of geographically distributed replica servers that deliver the content to end-users [1]. According to a Cisco Virtual Networking Index forecast [2], CDNs will carry about two-thirds of all Internet video traffic by 2020. This prediction has motivated CDN providers to provision Value Added Services (VASs), such as media management/video ad-insertion, in addition to their basic services, including video streaming. The building blocks of VASs often consist of a set of middle-boxes that needs to be deployed in the network and end-user traffic steers through them in a specific order.

Network Functions Virtualization (NFV) is an emerging technology that facilitates service provisioning by employing virtualization as a key technology [3]. It aims at decoupling the network functions (including middle-boxes) from the underlying hardware by defining them as standalone pieces of software called Virtual Network Functions (VNFs). In order to realize a VAS, such VNFs are chained in a specific order, forming a VNF Forwarding Graph (VNF-FG), a.k.a. a Service Function Chain (SFC). Fig.1 illustrates a use case [4] for ad-insertion as a CDN VAS. The required VNFs for ad-insertion VASs include i) a mixer for inserting advertisements, ii) a compressor for decreasing the video size/quality for devices with limited capabilities, and iii) a transcoder for video coding conversions. The VNF-FG for this VAS is formed such that for an end-user A, who wants to receive a low-quality video in AVI encoding, the video in mp4 format should pass through a mixer, a compressor and finally a transcoder before being delivered.

NFV enables the dynamic deployment and migration of VNF-FGs over NFV Infrastructure (NFVI) [5]. However, the location where VNFs are placed has a great impact on the total expenses of CDN providers. As discussed in [6], CDN VASs have a specific characteristic: one VNF-FG end-point is unknown prior to the placement. This end-point corresponds to the replica server that serves the content to the end-user. In concrete CDN scenarios, multiple replica servers may be capable of serving the end-user requested content according to their content availability. For example, in Fig.1 both replica servers X and Y can be assigned to end-user A. This dynamic assignment brings unique challenges to the problem as it impacts the optimal placement of VNFs in VNF-FG.

Although VNF-FG placement has attracted much research interest [7], to the best of our knowledge, very few research efforts focus on CDN VAS placement. Furthermore, such research works only propose an offline placement solution where the VNF-FGs are placed before service execution, i.e. before any request arrives and the video traffic steers into the network. This mode of placement is not adequate in cases where a change in service usage pattern occurs. An example of such situations is when a new group of end-users with new requirements on QoS subscribe to existing VASs. This calls for an online CDN VAS placement solution that efficiently reconfigures and adapts the system to the new changes. The reconfigurations are done in a way that some already-deployed VNFs could be reused or migrated to new locations, and some new VNFs may be instantiated. It should be noted that the existing online placement proposals do not deal with the specific case of VNF-FGs in CDN VASs, as mentioned above. In addition, none of them includes a multiple cost model that considers the costs of VNF migrations, new instantiations, hosting, and routing requests, together in one solution.

This paper deals with the problem of online VAS VNF-FG placement in CDNs. The objective is to optimally place the VNF-FGs so that the costs are minimized and the required QoS of all service requests are jointly satisfied. We consider a multiple cost model, called the reconfiguration cost that takes into account new VNF instantiations in addition to VNF hosting and migration. The routing of end-user service requests is also considered in this model. We model the problem as an Integer Linear Programming (ILP) optimization considering the reconfiguration costs. The problem is solved by using CPLEX optimization tool. The rest of the paper is organized as follows: Section II discusses the related work. Section III covers the system model and problem formulation. Section IV presents the evaluation results, and finally, in Section V we conclude the paper and explain the future work directions.

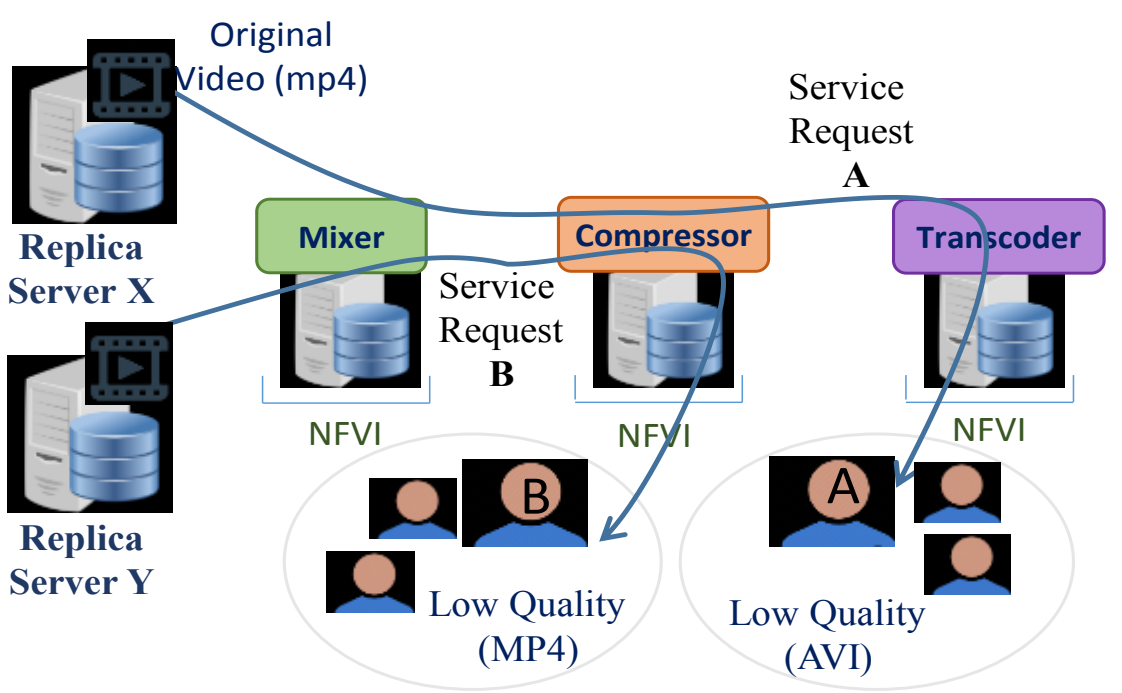

Fig. 1: ad-insertion CDN VAS use case.

## II. Related Work

We first focus on the works done so far on VNF-FG placement for CDN VASs. Next, we discuss the online VNF placement proposals in general.

Although VNF-FG placement has attracted the attention of researchers, very few of them have focused on CDN VAS placement. For instance, [8] tackles the CDN VAS placement problem by modeling the problem as an ILP formulation and proposes heuristics. The ILP focuses on minimizing cost while satisfying service request delay thresholds. The proposed heuristic includes two phases. In the first phase the VNFs are placed and chained, and in the second phase, the placement is improved by some additional constraints on VNF processing capacity. Reference [6] is another example of CDN VAS placement that proposes an ILP and a heuristics. In [6], a chain end-point is considered to be unknown before placement. The heuristics are based on a PageRank algorithm in which the requests with strict requirements are accommodated first. Both the above-mentioned studies focus on multiple objective costs for CDN VAS placement, including hosting, routing, and new VNF instantiations. Reference [9] is another related work on VAS placement in CDNs. It proposes a model for collaboration between ISPs and CDNs by defining SLAs in the form of service function chains. They propose CDN replica servers defined as VNFs and placed and deployed on NFVI provided by ISPs. In their proposed ILP, multiple costs are minimized, including routing and VNF hosting costs; however, their focus is on offline CDN VAS placement. It should be stressed again that all of these proposals focus on offline placement.

Research efforts on online VNF-FG placement are ongoing, however, they lack the required features to support VNF-FGs that do not have known end-points prior to placement. Apart from these shortcomings, they do not consider a multiple cost objective function, including both VNF migration and new instantiations. Some of the existing works consider a mono cost resolution, while others take into account multiple costs in their problem formulation.

References [10], [11], and [12] target a mono cost resolution for online VNF-FG placement. In [10], the authors address the problem of the optimal migration of VNFs in data centers to meet the computing and network resource constraints. Reference [11] focuses on the optimal placement of CDN replica servers defined as VNFs. Both works target minimizing the VNF migration cost in their ILP model, but without addressing new instantiation, hosting and routing costs. Reference [12] is another related work with mono cost resolution that does consider hosting costs. This work presents an ILP for online VNF placement and chaining, in environments with pure NFV settings and in hybrid environments containing physical and virtualized resources.

Among the works that take into account multiple costs, reference [13] proposes a solution for the online placement of VNFs in data centers in response to changing workloads. It presents an ILP formulation and heuristics and aims at jointly minimizing the migration, hosting and routing costs. Similarly, reference [14] jointly minimizes multiple costs, i.e. energy consumption and migration costs. Both these works consider multiple costs, however, they do not consider request routing or new VNF instantiations.

Unlike the existing literature, our work covers online VNF-FG placement for chains which have an end-point that is not known prior to placement. It proposes a multiple cost resolution model that includes VNF migrations and new instantiations, together. It not only targets the optimal placement of the VNFs and their chaining but also takes into account request routing.

## III. Online CDN VAS Placement Problem

### A. *Problem Description*

Assume that there is an ad-insertion VAS deployed in the network that is executing and satisfying the requests of existing end-users. For this VAS, a set of mixer, transcoder, and compressor VNFs are already deployed and chained in the network. Next, we assume a change in service usage occurs. Some examples are when a new group of end-users subscribes to an ad-insertion VAS, or when a new VAS is going to be introduced that shares a set of VNF types with already-deployed ad-insertion VAS. In such cases, some of the already-deployed VNFs can be reused, however, a reconfiguration may be needed to satisfy the QoS for all service requests.

It should be noted that there might be a set of service requests defined for each VAS. For example, in Fig. 1, two different service requests are defined for an ad-insertion VAS. Each service request requires a VNF-FG to be deployed and may cover several end-users (as shown in Fig. 1). Service requests are characterized by some service-related parameters. Examples of such parameters include the approximate location of end-users, the replica servers containing a popular video for end-users, the required VNF types and their order in the VNF-FG, the already-assigned VNF instances and routes, and the required QoS (e.g. a delay threshold tolerated by end-users belonging to that service request). The parameters of new service requests may be different from those of existing ones. For example, a new service request could need some new VNF types to be deployed, in addition to the already-deployed VNFs of existing services in the network.

The input to the problem includes the service-related parameters for existing services and new service requests, in addition to network-related parameters such as the location of already-deployed VNFs, and the available NFVI offered by replica servers, the cost, and the delays of links and servers. Given the above-mentioned inputs, the online CDN VAS placement problem consists of selecting the best content server, optimally placing the new required VNFs, migrating the existing ones, and connecting them to content servers and end-users, while the CDN provider costs are minimized and the QoS of all end-users are satisfied. The QoS is the service delay, calculated as the summation of the routing (i.e. communication) and video processing delays.

## B. *System Model*

We consider *N* as a set of network nodes, *S* as a set of replica servers, and *U* as a set of end-users, where $N = S \cup U$ . Given a network with *n* nodes, we represent the network topology with a matrix $B_{n\times n}$, where $B_{i,j}$ determines the bandwidth of the link between nodes *i* and *j*. Similarly, matrixes $C_{n\times n}$ and $D_{n\times n}$ determine the transmission cost and delay of the links between nodes. *K* denotes the set of VNF types defined in the system, such as mixer, transcoder, and compressor. Each VNF type $k \in K$ has a predefined license cost $L_k$ , processing capacity $P_k$, and resource requirement $R_k$. The set of available instances for VNF type k is delineated as $I_k$.

Each service request is indicated as *f*. The node indicating the end-user of service request *f*, the set of required VNF types for service request *f*, the first and last VNFs in VNF-FG, and the traffic load of service request *f* are respectively represented as $u_f$, $V^f$, $fst_f$, $lst_f$, and $T_f$. The maximum delay tolerated by service request *f* is denoted as $X_f$ .

Note that our model operates over two network snapshots: the *current* snapshot and the *new* one. The binary routing matrix $\tilde{P}^f_{n\times n}$ represents the links assigned to service request *f* in current snapshot, where $\tilde{p}^f_{i,j} \in \{0,1\}$ is equal to 1 if the link between nodes *i* and *j* is currently assigned to the service request *f*. Similarly, $\tilde{\tau}^s_{k,i}$ is a binary parameter, where $\tilde{\tau}^s_{k,i} \in \{0,1\}$ is equal to 1 if the instance *i* of VNF type *k* is instantiated on server *s* in the current snapshot. We define the following binary decision variables:

- $\gamma^f_s \in \{0,1\}$: specifies new content server assignment; if $\gamma^f_s$ is equal to 1, the server *s* is selected to serve the content for service request *f*.
- $\tau^s_{k,i} \in \{0,1\}$: specifies new VNF deployment topology; if $\tau^s_{k,i}$ is equal to 1, the VNF instance *i* of VNF type *k* is deployed on server *s*.
- $\lambda^f_{s,k,i} \in \{0,1\}$: specifies new VNF assignments; if $\lambda^f_{s,k,i}$ is equal to 1, the instance *i* of VNF type *k* instantiated on server *s* is assigned to service request *f*.

Table I. Input parameters and variables.

| | |
|---|---|
| **Inputs Parameters** | |
| $S$ | Set of servers |
| $U$ | Set of end-users |
| $N$ | Set of network nodes, $N = S \cup U$ |
| $K$ | Set of VNF types |
| $F$ | Set of service requests |
| $B_{i,j}$ | The bandwidth between nodes i and j, $i,j \in N$ |
| $C_{i,j}$ | The transmission cost between nodes i and j, $i,j \in N$ |
| $D_{i,j}$ | The transmission delay between nodes i and j, $i,j \in N$ |
| $L_k$ | License cost for VNF type *k*, $k \in K$ |
| $P_k$ | The processing capacity of VNF type *k*,(in traffic units), $k \in K$ |
| $R_k$ | The resource requirements of VNF type *k*, (in processing units), $k \in K$ |
| $I_k$ | Set of VNF instances associated to VNF type *k*, $k \in K$ |
| $\varphi^{s,t}_k$ | Cost of migrating VNF type *k* from server *s* to server t, $k \in K, s,t \in S$ |
| $M^s_k$ | The processing delay of VNF type *k* on server *s*, $k \in K,\ s \in S$ |
| $u_f$ | A node indicating the end-user of service request *f*, $u_f \in U, f \in F$ |
| $V^f$ | Set of required VNF types for service request *f*, $V^f \subset K$ |
| $fst_f$ | The first VNF in service chain of service request *f* , $fst_f \in V^f$ |
| $lst_f$ | The last VNF in service chain of service request *f* , $lst_f \in V^f$ |
| $T_f$ | Traffic units of service request *f* , $f \in F$ |
| $X_f$ | Maximum delay tolerated by service request *f* as per SLAs. |
| $\tilde{P}^f_{i,j}$ | 1, if currently the link between nodes i, j is assigned to service request *f*. |
| $\tilde{\tau}^s_{k,i}$ | 1, if currently instance *i* of VNF type *k* is instantiated on server *s*. |
| $\rho_s$ | Replica server's cost per unit, $s \in S$ |
| $G_s$ | Replica server's capacity (in processing resource units), $s \in S$ |
| $c^f_s$ | 1, if replica server ***s*** can be content server for service request *f*. |
| $\mu$ | Maximum node/link/VNF usage threshold |
| **Variables** | |
| $\gamma^f_s$ | Binary variable, indicating if server *s* is selected to serve content for service request *f* |
| $\tau^s_{k,i}$ | Binary variable, indicating if instance *i* of VNF type *k* is instantiated on server *s* |
| $\lambda^f_{s,k,i}$ | Binary variable, indicating if instance *i* of VNF type *k*, instantiated on server *s*, is assigned to service request *f* |
| $P^f_{i,j}$ | Binary variable, indicating if the link between nodes *i* and *j* will be assigned to service request *f* |

- $P^f_{i,j} \in \{0,1\}$: specifies new link assignments; if $P^f_{i,j}$ is equal to 1, the link between nodes *i* and *j* is assigned to service request *f*.

Table I summarizes the ILP parameters and variables.

## C. *Problem Formulation*

We formulate the problem of VAS placement in CDNs as an ILP formulation.

***Hosting cost ($\Delta C_{hst}$)-*** The hosting cost, is the differential cost of resources, between the two snapshots: new and current. Note that some resources might be released during reconfigurations that contribute in cost reductions.

$$\Delta C_{hst} = \sum_{s\in S}\sum_{k\in K}\sum_{i\in I_k} R_k.\rho_s.(\tau^s_{k,i} - \tilde{\tau}^s_{k,i}) \quad (1)$$

***Migration cost ($C_{mig}$)-*** As shown in Eq. (2), this is the total costs for migrating the already-deployed VNFs from one server to another.

$$C_{mig} = \sum_{s,t\in S}\sum_{k\in K}\sum_{i\in I_k} \varphi^{s,t}_k.\tilde{\tau}^s_{k,i}.\tau^t_{k,i} \quad (2)$$

Eq. (2) is not linear; to linearize it, we replace it with Eq. (2-1) and we consider Eqs. (2-2) to (2-4) as constraints.

$$C_{mig} = \sum_{s,t\in S}\sum_{k\in K}\sum_{i\in I_k} \varphi_k^{s,t}.X_{k,i}^{s,t} \quad , \quad X_{k,i}^{s,t} = \tilde{\tau}_{k,i}^{s}.\tau_{k,i}^{t} \qquad (2-1)$$

$$X_{k,i}^{s,t} \le \tilde{\tau}_{k,i}^{s} \qquad \forall s,t \in S, \forall i \in I_k, \forall k \in K \qquad (2-2)$$

$$X_{k,i}^{s,t} \le \tau_{k,i}^{t} \qquad \forall s,t \in S, \forall i \in I_k, \forall k \in K \qquad (2-3)$$

$$X_{k,i}^{s,t} \ge \tilde{\tau}_{k,i}^{s} + \tau_{k,i}^{t} - 1 \qquad \forall s,t \in S, \forall i \in I_k, \forall k \in K \qquad (2-4)$$

***VNF instantiation cost ($C_{inst}$)-*** This cost includes the total software license costs for new VNF instantiations, as shown in Eq. (3). It should be noted that the term $(\tau_{k,i}^{s} - \tilde{\tau}_{k,i}^{s})$ in this equation calculates the number of new VNF instantiations.

$$C_{inst} = \sum_{s\in S}\sum_{k\in K}\sum_{i\in I_k} L_k.(\tau_{k,i}^{s} - \tilde{\tau}_{k,i}^{s}) \qquad (3)$$

***Routing cost ($\Delta C_r$)-*** The routing cost, as shown in Eq. (4), is the differential cost of assigned links between two snapshots: new and current. This equation also includes the rerouting of currently existing service requests, that might happen in result of changing routes in reconfigurations.

$$\Delta C_r = \sum_{\forall f\in F}\sum_{s,t\in S} C_{s,t}.T_f.(P_{s,t}^{f} - \tilde{P}_{s,t}^{f}) \qquad (4)$$

Note that the routing ($\Delta C_r$) and hosting ($\Delta C_{hst}$) costs can be negative; i.e. the routing/hosting cost of VNF placement in the new snapshot is less than the current one.

***Objective-*** The objective is to minimize the total costs, as shown in Eq. (5).

$$Min(\Delta C_{hst} + C_{mig} + C_{inst} + \Delta C_r) \qquad (5)$$

***Constraints-*** The following constraints are considered in our model.

- **Content server constraints:** Eq. (6) ensures that only one content server is selected to serve the service request *f*, and Eq. (7) ensures that the content server is selected from the capable replica servers.

$$\sum_{\forall s\in S} \gamma_s^f = 1 \qquad \forall f \in F \qquad (6)$$

$$\gamma_s^f.c_s^f = \gamma_s^f \qquad \forall s \in S, \forall f \in F \qquad (7)$$

- **VNF assignment constraints:** Eq. (8) ensures that only one instance of each required VNF type is assigned to service request *f*. Eq. (9) ensures that the assigned VNF instances are already deployed in the network.

$$\sum_{\forall s\in S}\sum_{\forall i\in I_k} \lambda_{s,k,i}^{f} = 1 \qquad \forall k \in V^f, \forall f \in F \qquad (8)$$

$$\gamma_s^f.c_s^f = \gamma_s^f \qquad \forall s \in S, \forall f \in F \qquad (9)$$

- **VNF deployment constraint:** Eq. (10) ensures that at least one instance of each required VNF type is deployed and Eq. (11) ensures that each VNF instance is deployed not more than once in the network.

$$\sum_{\forall s\in S}\sum_{\forall i\in I_k} \tau_{k,i}^{s} \ge 1 \qquad \forall k \in K \qquad (10)$$

$$\sum_{\forall s\in S} \tau_{k,i}^{s} \le 1 \qquad \forall k \in K, \forall i \in I_k \qquad (11)$$

- **Server capacity constraint:** Eq. (12) ensures that the replica servers hosing VNFs are not overloaded.

$$\sum_{k\in K}\sum_{i\in I_k} R_k.\tau_{k,i}^{s} \le \mu.G_s \qquad \forall s \in S \qquad (12)$$

- **VNF capacity constraint:** Eq. (13) ensures that the VNFs are not overloaded.

$$\sum_{f\in F} T_f.\lambda_{s,k,i}^{f} \le \mu.P_k \qquad \forall k \in K, \forall i \in I_k, \forall s \in S \qquad (13)$$

- **Link capacity constraint:** Eq. (14) ensures that the links assigned to service requests are not overloaded.

$$\sum_{f\in F} T_f.P_{s,t}^{f} \le \mu.B_{s,t} \qquad \forall s,t \in N \qquad (14)$$

- **Link assignment constraints:** Eq. (15) ensures that a link is assigned between a selected content server and the first VNF in the chain of service request *f*. Eq. (16) ensures that a link is assigned between each pair of VNFs in the service request *f*. We assume that VNFs are traversed in the same numerical order as considered in [15], therefore, in Eq. (16), $VNF_{m+1}$ is the next VNF after $VNF_m$ in a set of required VNFs for service request *f*. Eq. (17) ensures that a link is assigned between the last VNF in the chain and the end-user of service request *f*.

$$\gamma_s^f.\lambda_{t,fst_f,i}^{f} \le P_{s,t}^{f} \qquad \forall s,t \in S, \forall f \in F, i \in I_{fst_f} \qquad (15)$$

$$\lambda_{s,m,i}^{f}.\lambda_{t,m+1,j}^{f} \le P_{s,t}^{f} \qquad \forall s,t \in S, \forall f \in F, m \in V^f, i \in I_m, j \in I_{m+1} \qquad (16)$$

$$\lambda_{s,lst_f,i}^{f} = P_{s,u_f}^{f} \qquad \forall s \in S, \forall f \in F, i \in I_{lst_f} \qquad (17)$$

Eq. (15) is non-linear, however, it can be linearized by replacing it with linear equations (15-2) to (15-5):

$$\gamma_s^f.\lambda_{t,fst_f,i}^{f} = M_{s,t,fst_f,i}^{f} \qquad (15-1)$$

$$M_{s,t,fst_f,i}^{f} \le P_{s,t}^{f} \qquad \forall s,t \in S, \forall f \in F, \forall i \in I_{fst_f} \qquad (15-2)$$

$$M_{s,t,fst_f,i}^{f} \le \gamma_s^f \qquad \forall s,t \in S, \forall f \in F, \forall i \in I_{fst_f} \qquad (15-3)$$

$$M_{s,t,fst_f,i}^{f} \le \lambda_{t,fst_f,i}^{f} \qquad \forall s,t \in S, \forall f \in F, \forall i \in I_{fst_f} \qquad (15-4)$$

$$M_{s,t,fst_f,i}^{f} \ge \lambda_{t,fst_f,i}^{f} + \gamma_s^f - 1 \qquad \forall s,t \in S, \forall f \in F, \forall i \in I_{fst_f} \qquad (15-5)$$

Similarly, non-linear Eq. (16) can be replaced with linear Eqs. (16-2) to (16-5):

$$Q_{s,t,m,m+1,i,j}^{f} = \lambda_{s,m,i}^{f}.\lambda_{t,m+1,j}^{f} \qquad (16-1)$$

$$Q_{s,t,m,m+1,i,j}^{f} \le P_{s,t}^{f} \qquad \forall s,t \in S, \forall f \in F, m \in V^f, i \in I_m, j \in I_{m+1} \qquad (16-2)$$

$$Q_{s,t,m,m+1,i,j}^{f} \le \lambda_{s,m,i}^{f} \qquad \forall s,t \in S, \forall f \in F, m \in V^f, i \in I_m, j \in I_{m+1} \qquad (16-3)$$

$$Q_{s,t,m,m+1,i,j}^{f} \le \lambda_{t,m+1,j}^{f} \qquad \forall s,t \in S, \forall f \in F, m \in V^f, i \in I_m, j \in I_{m+1} \qquad (16-4)$$

$$Q_{s,t,m,m+1,i,j}^{f} \ge \lambda_{s,m,i}^{f} + \lambda_{t,m+1,j}^{f} - 1 \qquad \forall s,t \in S, \forall f \in F, m \in V^f, i \in I_m, j \in I_{m+1} \qquad (16-5)$$

- **QoS satisfaction constraint:** Eq. (18) ensures that the required QoS (in terms of service delay) for each service request is satisfied. The delay is calculated as the sum of the transmission delay and the video processing delay.

$$\sum_{\forall s,t\in N} T_f.D_{s,t}.P^f_{s,t} + \sum_{\substack{\forall k\in V^f \\ \forall i\in I_k \\ \forall s\in S}} T_f.M^s_k.\lambda^f_{s,k,i} \le X_f \quad \forall\in F \tag{18}$$

# I. PERFORMANCE EVALUATION

In this section, we describe our simulation setup and evaluation scenario and then we present our results.

## A. Simulation Setup and Evaluation Scenario

We consider a network with 12 nodes: 6 replica servers and 6 groups of end-users. We assume logical links exist between each pair of nodes, that each link has 10 Gbps of bandwidth capacity [16] and that the bandwidth cost for each link is randomly given between 0.115[1] and 0.09[2] \$/GB. We map each node location to a city in the USA and determine the delay between nodes using WonderNetwork[3], which provides the real-time hourly delay information between various pairs of locations. We consider 6 service requests, each with a random delay threshold ranging from 1800-2000 ms [17]. Each service request requires 1 to 3 VNFs, given randomly. At least 3 replica servers are selected randomly as the content server for the end-users of each service request. We consider a license cost of \$100 [6]. We assume each VNF uses a medium OpenStack VM, with 2 vCPUs, a 40GB disk and 4GB of memory for execution [16]. The cost for migrating a VNF from an original server to a new one is calculated, as the bandwidth cost of migrating the VNF (the sum of its disk and memory size), based on references [13] and [16]. Therefore, in our simulations, VNF migration cost is equal to 40+4 GB multiplied by the cost of the link between those two servers. We consider each server can host a maximum 4 VNFs [13] with the cost of server usage set as 5\$/vCPU [6]. Table II summarizes the simulation parameters.

Multiple scenarios are implemented with different numbers of existing (already deployed) and newly-arrived service requests, as detailed in Table III. We vary the number of existing and new service requests with the objective of evaluating their impact on the online VNF placement problem. For each scenario, we implement two cases: i) first case implements our proposed online placement solution and ii) second case implements the approach where the required VNFs of new service requests are deployed from scratch, i.e. the already-deployed VNFs are not reused for new service requests. This helps to obtain some insights about the advantage of reusing the already-deployed VNFs. The proposed online placement ILP model is implemented and solved in IBM CPLEX 12.8.0.0.

## B. Evaluation Results

The total reconfiguration costs are shown in Fig. 2 (a). As can be observed, there are notable differences between the two cases; clearly, when VNFs are not reused there are higher reconfiguration costs compared to the online placement method. Comparing the total cost of online placement for three scenarios, Fig. 2 (a) shows that as the number of new service requests in each scenario decreases (compared to the number of existing service requests), the reconfiguration cost is reduced as well. For example, for scenario 1 with 2 existing and 4 new requests, the reconfiguration cost of online placement is \$330.03 and is reduced to \$110.182 in the third scenario with 4 existing and 2 new service requests. This is because when the number of new service requests is small, they can be easily accommodated by existing VNFs, by reusing existing VNFs. Furthermore, when we move from scenario 1 to 3, as the number of existing service requests increases, more cost reduction is achieved in online placement in comparison to the case when VNFs are not reused.

Table II. Simulation Parameters.

| **Parameter** | **ILP Notation** | **Value** |
|---|---|---|
| Number of servers | - | 6 |
| Number of end-user groups | - | 6 |
| Link bandwidth capacity (Gb/s) | $B_{i,j}$ | 10 |
| Link bandwidth cost (\$/GB) | $C_{i,j}$ | Rand [0.115, 0.09] |
| Link delay (ms) | $D_{i,j}$ | 4-50 |
| Service request delay threshold (ms) | $X_f$ | Rand [1800-2000] |
| VNFs in each service request | $V^f$ | Rand [1-3] |
| VNF license cost (\$) | $L_k$ | 100 |
| VNF resource requirements (vCPU) | $R_k$ | 2 |
| VNF processing delay (ms) | $M^s_k$ | 20 |
| Replica servers' capacity (vCPU) | $G_s$ | 8 |
| Replica servers' cost (\$/vCPU) | $\rho_s$ | 5 |
| Traffic units of service request *f* (GB) | $T_f$ | 1 |
| Maximum usage threshold | $\mu$ | 1 |

Table III. Evaluation Scenarios

| Scenario / Service Request | Scenario 1 | Scenario 2 | Scenario 3 |
|---|---|---|---|
| Existing/ already deployed | 2 | 3 | 4 |
| Newly introduced | 4 | 3 | 2 |

Figures 2(b, c, and d) further show the details of reconfiguration cost, i.e. VNF hosting, VNF instantiation and request routing costs, respectively, for both online placement and where VNFs are not reused. As can be observed, there are notable differences between the two cases, highlighting the advantages of the proposed online placement approach in which already-deployed VNFs are reused, for a lower cost. It also can be observed that VNF instantiations and hosting costs have more significant impact on the total reconfiguration costs, compared to routing costs, due to the fact that the VNF license and hosting expenses are much more than bandwidth expenses. It should be noted that the migration costs are not shown in this figure. The measured VNF migration costs for all scenarios were zero, which indicates that no migration has occurred during executions. This shows the significant overhead of live VNF migrations and is due to the fact that in live migration [18], the whole VM that is hosting the VNF is migrated to another location, imposing high transportation costs. The scenarios were run on a server with 2×12-Core 2.20 GHz Intel Xeon E5-2650 v4 CPUs with 128GB of memory. Note that as we move from scenario 3 to 1, the execution time of our proposed online VNF placement changes on the order of seconds to hours.

---

[1] https://aws.amazon.com/govcloud-us/pricing/data-transfer/
[2] https://www.ibm.com/cloud/bandwidth
[3] https://wondernetwork.com/

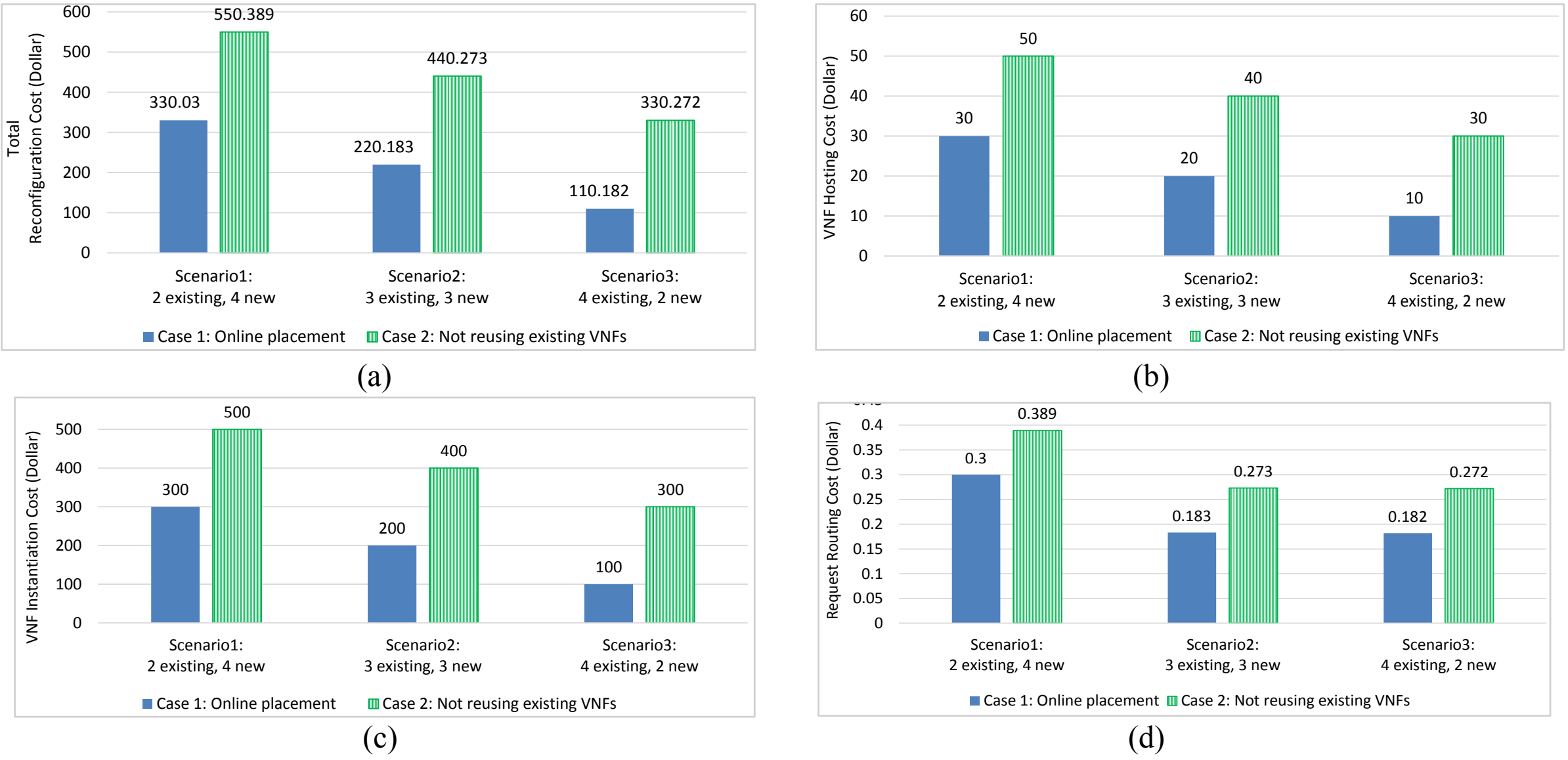

Fig.2. Evaluation results. a) Total reconfiguration cost, b) VNF hosting cost, c) VNF instantiation cost, d) Request routing cost.

## II. Conclusion

This paper studies the problem of online VNF-FG placement for VASs in CDNs, taking into account the reuse of already-deployed VNFs. It proposes a cost model including new VNF instantiations, migration, hosting, and routing costs. The objective is to optimally place the VNFs so that reconfiguration costs are minimized while respecting the QoS of all service requests. The problem is formulated as an ILP and evaluated in comparison with the case where already-deployed VNFs are not reused for new service requests. The results show that the proposed online placement method outperforms the approach where VNFs are not reused. However, the long execution times for large scale scenarios indicate the need for adequate online placement algorithms, which is planned as future work. Furthermore, the high cost of live VNF migration calls for investigations into alternative migration methods such as VNF state migration.

## Acknowledgment

This work was supported in part by Ericsson Canada and the Natural Science and Engineering Council of Canada (NSERC).

## References

[1] M. Wang *et al.*, "An Overview of Cloud Based Content Delivery Networks: Research Dimensions and State-of-the-Art," in *Transactions on Large-Scale Data- and Knowledge-Centered Systems XX*, A. Hameurlain, J. Küng, R. Wagner, S. Sakr, L. Wang, and A. Zomaya, Eds. Springer Berlin Heidelberg, 2015, pp. 131–158.

[2] "Cisco Visual Networking Index: Forecast and Methodology, 2016–2021," *Cisco*. [Online]. Available: https://www.cisco.com/c/en/us/solutions/collateral/service-provider/visual-networking-index-vni/complete-white-paper-c11-481360.html. [Accessed: 28-Nov-2017].

[3] B. Han, V. Gopalakrishnan, L. Ji, and S. Lee, "Network function virtualization: Challenges and opportunities for innovations," *IEEE Commun. Mag.*, vol. 53, no. 2, pp. 90–97, Feb. 2015.

[4] N. T. Jahromi *et al.*, "NFV and SDN-based cost-efficient and agile value-added video services provisioning in content delivery networks," in *2017 14th IEEE CCNC*, 2017, pp. 671–677.

[5] Network Functions Virtualization (NFV): Architectural framework G ETSI - ETSI GS NFV, 2014.

[6] M. Dieye *et al.*, "CPVNF:Cost-efficient Proactive VNF Placement and Chaining for Value-Added Services in Content Delivery Networks," *IEEE Trans. Netw. Serv. Manag.*, vol. PP, no. 99, pp. 1–1, 2018.

[7] J. G. Herrera and J. F. Botero, "Resource Allocation in NFV: A Comprehensive Survey," *IEEE Trans. Netw. Serv. Manag.*, vol. 13, no. 3, pp. 518–532, Sep. 2016.

[8] S. Ahvar *et al.*, "PCPV: Pattern-based Cost-efficient Proactive VNF Placement and Chaining for Value-Added Services in Content Delivery Networks," accepted for publication in the IEEE Netsoft, Canada, 2018.

[9] N. Herbaut, D. Negru, D. Dietrich, and P. Papadimitriou, "Service chain modeling and embedding for NFV-based content delivery," in *2017 IEEE Int. Conf. on Com. (ICC)*, 2017, pp. 1–7.

[10] J. Xia, Z. Cai, and M. Xu, "Optimized Virtual Network Functions Migration for NFV," in *2016 IEEE 22nd International Conference on Parallel and Distributed Systems (ICPADS)*, 2016, pp. 340–346.

[11] "OPAC: An optimal placement algorithm for virtual CDN - ScienceDirect." [Online]. Available: https://www.sciencedirect.com/science/article/pii/S1389128617301391. [Accessed: 12-Mar-2018].

[12] "VNF-P: A model for efficient placement of virtualized network functions - IEEE Conference Publication." [Online]. Available: http://ieeexplore.ieee.org/abstract/document/7014205/. [Accessed: 12-Mar-2018].

[13] M. Ghaznavi, A. Khan, N. Shahriar, K. Alsubhi, R. Ahmed, and R. Boutaba, "Elastic virtual network function placement," in *2015 IEEE 4th International Conference on Cloud Networking (CloudNet)*, 2015, pp. 255–260.

[14] V. Eramo, E. Miucci, M. Ammar, and F. G. Lavacca, "An Approach for Service Function Chain Routing and Virtual Function Network Instance Migration in Network Function Virtualization Architectures," *IEEEACM Trans. Netw.*, vol. 25, no. 4, pp. 2008–2025, Aug. 2017.

[15] D. Bhamare, M. Samaka, A. Erbad, R. Jain, L. Gupta, and H. A. Chan, "Optimal virtual network function placement in multi-cloud service function chaining architecture," *Comput. Commun.*, vol. 102, pp. 1–16, Apr. 2017.

[16] M. Abu-Lebdeh, D. Naboulsi, R. Glitho, and C. W. Tchouati, "On the Placement of VNF Managers in Large-Scale and Distributed NFV Systems," IEEE Trans. Netw. Service Manag. , vol. PP, pp. 1–1, 2017.

[17] R. A. Cacheda, D. C. Garc´ıa, A. Cuevas, and F. Casta˜no, "QoS requirements for multimedia services," *Resour. Manag. Satell. Netw. Springer Boston MA*, pp. 67–94, Jan. 2007.

[18] C. Mouradian *et al.*, "Network functions virtualization architecture for gateways for virtualized wireless sensor and actuator networks," *IEEE Netw.*, vol. 30, no. 3, pp. 72–80, May 2016.